\documentstyle[multicol,aps,prb,epsf]{revtex}

\begin{document}
\pagestyle{empty}

\newcommand{\bc}{\begin{center}}
\newcommand{\ec}{\end{center}}
\newcommand{\be}{\begin{equation}}
\newcommand{\ee}{\end{equation}}
\newcommand{\beqn}{\begin{eqnarray}}
\newcommand{\eeqn}{\end{eqnarray}}

\begin{multicols}{2}
\narrowtext
\parskip=0cm

\noindent
{\large\bf Comment on "Aging Effects in a Lennard-Jones Glass"}
\smallskip


In a recent Letter Kob and Barrat \cite{kob} reported results of
molecular dynamics simulations for the off-equilibrium dynamics in a
binary Lennard-Jones (LJ) glass. The main conclusions of their work
was 1) they find aging in this glassy systems and 2) that they find a
{\it simple} aging scenario close to a $t/t_w$ scaling, which is very
reminiscent of comparable studies in spin glasses. In this comment we
would like to emphasize that a different aging scenario, known under
the name {\it activated dynamics} scaling, is much more appropriate
for the system under consideration than the one proposed by Kob and
Barrat \cite{kob}.

For this reason we repeated the simulation by Kob and Barrat, using
exactly the same potential (Lennard-Jones for a binary mixture), the
same parameters (same diameters, mixture, density and temperatures)
and the same quenching procedure ($T_i$=5, $T_f$=0.4) however with
much larger systems ($32768 = 32^3$ particles) and similar times
($2\cdot 10^6$ time steps, 1 time step corresponding to 0.01
LJ-units). The aging properties of the system manifest themselves in
the two-time autocorrelation function
\be\label{aging_function}
  C_q(t+t_w, t_w) = \frac{1}{N}
  \sum_{i} e^{i \cdot q \cdot \left[r_i(t+t_w) - r_i(t_w)\right] }\;,
\ee
where $r_i(t)$ is the position of particle $i$ at time $t$ and the
absolute value of $q$ corresponds to the first maximum in the
structure function. We choose 100 randomly distributed vectors and
averaged $C_q$ over these vectors. The function (\ref{aging_function})
was evaluated after every 10 time steps and $5^n$ measurements were
averaged over to improve statistics.  We convinced ourselves that
different quenching procedures with identical initial and final
temperatures, $T_i$ and $T_f$, lead to the same scaling behavior.

In [1] it has been suggested that $C_q(t+t_w, t_w)$ obeys
\be
C_q(t+t_w, t_w) \sim \tilde{c}(t/t_r)
\label{simple}
\ee
with a relaxation time $t_r\propto t_w^\alpha$. We checked this {\it
  Ansatz} for our data and display the result in the inset of Fig. 1,
surprisingly we find an exponent $\alpha\sim1.1$, very close to one
(corresponding to simple $t/t_w$ scaling) but different from the one
$\alpha=0.88$ reported in [1]. The data collapse in the asymptotic
regime is not at all satisfying, the data for different waiting times
conincide exactly only for $C_q=0.45$. For this reason we tried
another aging scenario, proposed in the context of spin glasses by
Fisher and Huse \cite{fh}, which we call the {\it activated dynamics}:
\be
C_q(t+t_w, t_w) \sim \tilde{C}\bigl\{\ln((t+t_w)/\tau)/\ln(t_w/\tau)\bigr\}
\label{activated}
\ee
where $\tau$ is a fit-parameter and plays the role of an effective
microscopic time scale. Fig.1 we show the scaling plot for such a
scenario, which gives a much better data collapse in the asymptotic
regime $t\ge t_w$.

\begin{figure}
\epsfxsize=\columnwidth\epsfbox{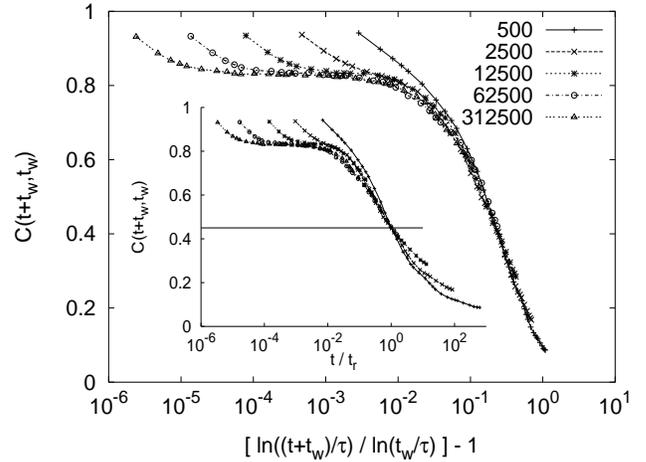}
\vspace{0.3cm}
\caption{Activated dynamics scaling plot according to eq.(\ref{activated}) with
  $\tau=0.5$. Note that we shifted the scaling variable by 1 to the
  left to have a better resolution of the crossover region.
  The inset shows a scaling plot of our data according to
  the scenario proposed by Kob and Barrat [1], see
  eq.(\protect{\ref{simple}}), the full line corresponds to
  $C_q=0.45$. The relaxation times are $t_r$=1450, 10600, 70000,
  600000 and 3000000 for $t_w$=500, 2500, 12500, 62500 and 312500,
  roughly a dependence $t_r\propto t_w^{1.1}$.
\label{fig}
}
\end{figure}

The origin of such an activated dynamics scaling in spin glass
phenomenology \cite{fh} is simply a logarithmically slow coarsening
process $\xi(t)\sim\ln(t)^a$, where $\xi(t)$ is a time dependent
spatial correlation length and $a$ some exponent. This plus the
observation that in coarsening dynamics the two time correlation
function $C_q(t+t_w, t_w)$ should depend on the ration of the two
length scale $\xi(t_w)/\xi(t+t_w)$ alone yields the aging behavior
(\ref{activated}).

Three things are worth being noted: 1) In the context, in which
(\ref{activated}) was first suggested, namely the 3d EA spin glass,
this form does not seem to work \cite{sg_aging}. 2) Only very recently
a growing length scale has been observed in the very same model we are
considering here \cite{parisi}. 3) An even better data collapse can be
obtained by plotting $C_q(t+t_w, t_w)$ versus $\ln(t)/\ln(t_r)$, with
a relaxation time $t_r$ individually chosen for each waiting time
$t_w$. Here it turns out that $t_r(t_w)$ grows faster than with a power law.

To conclude we have shown that the aging behavior of a Lennard-Jones
glass is more appropriately described by an activated dynamics scaling
rather than simple aging, as claimed by Kob and Barrat in [1].

\bigskip
\noindent
Uwe M\"ussel and Heiko Rieger

{\small

HLRZ c/oForschungszentrum J\"ulich

52425 J\"ulich, Germany

}
\bigskip
\noindent
Date: 18 February 1998


\end{multicols}

\end{document}